\documentclass[manuscript]{acmart}
\usepackage{tabularx}
\usepackage{booktabs}
\AtBeginDocument{%
  }
\usepackage{todonotes}

\setcopyright{acmlicensed}
\copyrightyear{}
\acmYear{}
\acmDOI{XXXXXXX.XXXXXXX}
\acmConference[Conference acronym 'XX]{Make sure to enter the correct
  conference title from your rights confirmation email}{date year}{location}
\acmISBN{}

\usepackage[most]{tcolorbox}
\definecolor{chiBlue}{RGB}{0, 84, 147}

\tcbset{
  sigchibox/.style={
    enhanced,
    breakable,               
    sharp corners,
    boxrule=1pt,
    colframe=chiBlue,
    colback=chiBlue!10,
    left=2mm, right=2mm, top=1mm, bottom=1mm,
    fonttitle=\bfseries\small,
    title=Design Implication,
    width=\linewidth,        
    attach boxed title to top left={yshift=-2mm,xshift=2mm},
    boxed title style={
      colback=chiBlue,
      colframe=chiBlue,
      sharp corners
    }
  }
}

\newtcolorbox{designimp}[1][]{
  sigchibox,
  title={#1}              
}

\begin{document}

\title{Exploring Self-Tracking Practices of Older Adults with CVD to Inform the Design of LLM-Enabled Health Data Sensemaking}

\author{Duosi Dai}
\authornote{The work was done when the author worked as an ERASMUS-Trainee in Ludwig Boltzmann Institute for Digital Health and Prevention.}
\email{duosi@cs.au.dk}
\affiliation{%
  \institution{Aarhus University}
  \city{Aarhus}
  \state{Aarhus}
  \country{Denmark}
}

\author{Pavithren V S Pakianathan}
\email{pavithren.pakianathan@dhp.lbg.ac.at}
\affiliation{%
  \institution{Ludwig Boltzmann Institute for Digital Health and Prevention}
  \city{Salzburg}
  \country{Austria}
}

\author{Gunnar Treff}
\affiliation{%
  \institution{Ludwig Boltzmann Institute for Digital Health and Prevention}
  \city{Salzburg}
  \country{Austria}
}

\author{Mahdi Sareban}
\affiliation{%
  \institution{Ludwig Boltzmann Institute for Digital Health and Prevention}
  \city{Salzburg}
  \country{Austria}
}

\author{Jan David Smeddinck}
\affiliation{%
  \institution{Ludwig Boltzmann Institute for Digital Health and Prevention}
  \city{Salzburg}
  \country{Austria}
}

\author{Sanna Kuoppamäki}
\email{Department of Biomedical Engineering and Health Systems}
\affiliation{%
  \institution{KTH Royal Institute of Technology}
  \city{Stockholm}
  \country{Sweden}
}

\renewcommand{\shortauthors}{Dai et al.}

\begin{abstract}
Wearables and mobile health applications are increasingly adopted for self-management of chronic illnesses; yet the data feels overwhelming for older adults with cardiovascular disease (CVD). This study explores how they make sense of self-tracked data and identifies design opportunities for Large Language Model (LLM)-enabled support. We conducted a seven-day diary study and follow-up interviews with eight CVD patients aged 64–82. We identified six themes: navigating emotional complexity, owning health narratives, prioritizing bodily sensations, selective engagement with health metrics, negotiating socio-technical dynamics of sharing, and cautious optimism toward AI. Findings highlight self-tracking is affective, interpretive, and socially situated. We outline design directions for LLM-enabled data sensemaking systems: supporting emotional engagement, reinforcing patient agency, acknowledging embodied experiences, and prompting dialogue in clinical and social contexts. To support safety, expert-in-the-loop mechanisms are essential. These directions articulate how LLMs can help translate data into narratives and carry implications for human-data interaction and behavior-change support.
\end{abstract}

\ccsdesc[500]{Human-centered computing~Empirical studies in HCI}
\ccsdesc[300]{Human-centered computing~Interaction design theory, concepts and paradigms}
\ccsdesc[500]{Applied computing~Health informatics}

\keywords{Wearable technology, Older adults, Cardiovascular disease, Self-tracking, Data sensemaking, Digital health, Personal informatics}



\maketitle

\section{Introduction}
Self-tracking technologies such as wearable devices and mobile health apps have become widely integrated into everyday life, offering new ways for individuals to manage their health through collection of personalized data \cite{feng_how_2021}. These tools enable individuals to track physical activity, sleep and heart rate with high accuracy due to continuous improvements in sensor technology \cite{doherty_keeping_2024}. Individuals must engage in a process of data interpretation, commonly known as data sensemaking, where they reflect and act upon data they collected \cite{li_stage-based_2010, coskun_data_2023}. This process is particularly nuanced and challenging for older adults, for whom self-tracking is often linked to chronic illness management, emotional reassurance, and maintaining independence \cite{jaana_comparison_2020,nurain_i_2023,kononova_use_2019}.

Our work explores the self-tracking practices of older adults with CVD, with the aim of informing the potential design of LLM-based tools that support their sensemaking of self-tracked data \cite{tadas_using_2023}. It looks into the strategies and challenges this group faces when using wearables and health apps for tracking physical activity. It also explores how LLMs, with their ability to turn complex data into simple and context-rich language, can match the tracking preferences of older CVD patients by offering feedback that is meaningful, personal, and emotionally supportive. This is important to research since older adults often have their own ways of understanding and reacting to their health data \cite{wang_redefining_2024, vargemidis_performance_2023}.

Older adults managing cardiovascular disease (CVD) represent a particularly important and yet underserved population in this context \cite{zhou_aging_2022}. While regular physical activity is essential for CVD secondary prevention and management, sustaining motivation and tracking progress after structured rehabilitation programmes concluded is difficult \cite{graham_systematic_2020}. For this group, wearable devices and health apps offer the potential to support long-term self-care \cite{jurgen_vogel_digitally_2017, tadas_using_2023}. However, studies show that the tools are often inaccessible or difficult to use due to age-related cognitive changes, unfamiliar design language, and a lack of meaningful or personalized feedback \cite{jaana_comparison_2020, domingos_usability_2022,harrington_designing_2018}. Current wearables and mobile health applications are typically designed with younger or general populations in mind, and they frequently fail to reflect the physical, emotional, and contextual needs of health older adults and those living with chronic illnesses \cite{wang_redefining_2024,bostrom_mobile_2020}. Instead of offering clarity, the resulting data might become overwhelming or bring worry and anxiety to older adults with CVD \cite{ancker_you_2015}. At the same time, some countries have started reimbursing digital health applications for chronic conditions. This indicates a growing recognition of such tools as part of formal care pathways \cite{van2023digital}. This contrast highlights the need for solutions that are both clinically recognized and more usable for older adults managing chronic conditions. 

Older adults living with CVD were chosen as the focus of this work because CVD is both highly prevalent in this age group \cite{zhou_aging_2022} and closely linked to metrics already monitored through self-tracking technologies, such as heart rate, ECG, oxygen level, and physical activity \cite{ferguson_wearables_2021,krishnaswami_gerotechnology_2020,bostrom_mobile_2020}. At the same time, previous studies show that CVD patients often experience confusion, anxiety, or emotional stress when data or alerts are presented without sufficient explanation \cite{rosman_when_2020,varma2024promises}, especially in later stages of rehabilitation or long-term condition management when they must rely on their own interpretation. Older adults' interpretation of self-tracked data is shaped by their needs, preferences, and lived experiences \cite{vargemidis_performance_2023, brickwood_older_2020, ferguson_wearables_2021, wang_redefining_2024}. They combine digital feedback with body awareness and emotions, rather than passively accepting data. Yet, many self-tracking systems provide general tracking but lack the responsiveness and flexibility to support these interpretive practices \cite{li_understanding_2011,fritz_persuasive_2014,chung_more_2015,kononova_use_2019}. As a result, older adults with CVD present both practical access to rich self-tracking data and a pressing need for more personalized and accessible sensemaking tools.

\subsection{Research Questions}
The recent rise of generative AI, particularly through Large Language Models (LLMs), presents new opportunities for personalized support in digital health. LLM-based tools have shown potential in delivering feedback in natural language, supporting health literacy, and engaging individuals in reflective dialogue \cite{jorke2024supporting,stromel2024narrating,fang_physiollm_2024}. However, most existing systems are designed for general adult populations and fail to address the emotional tone, lived experiences, and contextual demands of older adults with chronic illness \cite{kaliappan_exploring_2024,enam_artificial_2025}. This leaves a gap in our understanding of how to support this group in sustaining physical activity and deriving personal meaning from their self-tracked health data. Our work addresses these gaps by investigating two key questions:

\begin{itemize}
    \item \textbf{RQ1}: How do strategies, motivations, and perceived barriers shape the ways CVD patients aged 60 and above make sense of their health data through self-tracking technologies?
    
    \item \textbf{RQ2}: What design opportunities for LLM-enabled interactive systems can help better scaffold support for older adults with CVD in interpreting and reflecting on their self-tracked health data? 
\end{itemize}

Through a combination of user studies and design direction discussions, this work aims to explore the engagement of older CVD patients in self-tracking to inform opportunities and design directions for LLM-enabled tools in supporting their personalized reflection and ongoing health management.

\subsection{Contributions}
This work makes the following contributions to HCI and digital health:

\begin{itemize}
    \item An empirical study with eight older adults living with CVD to understand how they interpret data from wearables and mHealth apps. Through a seven-day diary study and in-depth interviews, we offer insights into how this specific user group makes sense of their health data. 


    \item Design directions for self-tracking technologies based on LLMs to \emph{support data sensemaking}, \emph{including emotionally responsive feedback}, \emph{reinforcement of agency and autonomy}, \emph{acknowledgement of embodied experiences}, and \emph{positioning AI as a mediator within social and clinical contexts} rather than a replacement for human judgement. 
\end{itemize}

\section{Background}
\subsection{Self-tracking and data sensemaking with wearables}

Self-tracking technologies, such as wearables and mHealth apps, have become part of people’s everyday routines, not only for optimizing performance or fitness, but also for maintaining health and chronic disease self-care \cite{feng_how_2021}. This reflects a broader cultural shift in which bodies and health are interpreted through the lens of continuous data streams \cite{Ruckenstein2017Datafication}. Individuals are positioned as both subjects and agents of data production \cite{li_stage-based_2010,lupton2016quantified}. Central to these practices is 'data sensemaking', which is the ongoing interpretation, contextualization, and negotiation of meaning from self-tracked numbers \cite{coskun_data_2023, Fiske2019DataWork}. This resonates with Mortier et al.’s concept of human-data interaction governed by legibility, agency, and negotiability \cite{mortier2014human}.

However, self-tracking is rarely a straightforward path from data to behavior change. It is, instead, fragmented and shaped by personal goals, rhythms, and anxieties \cite{rooksby_personal_2014,lupton2014self,li_stage-based_2010,dehghani2018exploring}, as well as by an ever-changing landscape of developing technologies and a lack of data harmonization. The emotional dimension of data is especially important in health and aging. Data can reassure but also overwhelm, particularly when information is complex, fragmented across platforms, or perceived as externally imposed \cite{Kaziunas2017CaringData,lupton2014self,jones_dealing_2018}. Without adequate contextual support, self-tracking may induce confusion or feelings of inadequacy rather than empowerment \cite{fritz_persuasive_2014}. Recent work calls for tools that move beyond purely data-focused dashboards toward systems that support interpretation, manage uncertainty, and respond to emotional reactions. Rather than presenting numbers only, these tools aim to engage users in dialogue and democratize access to meaningful data interpretation \cite{puussaar2020democratising}. 

\subsection{Self-tracking among cardiovascular disease patients}

Supporting monitoring outside traditional health systems not only helps chronic patients manage daily needs but can also improve system-level efficiency, reducing healthcare costs and contributing to quality-adjusted life years (QALYs) \cite{drummond2015methods}. In the context of cardiovascular disease -- the largest cause of death worldwide -- qualitative evidence \cite{tadas2020barriers} suggests that many cardiac patients value self-tracking and monitoring because it strengthens both personal and in-the-moment understanding of their condition. By making health and contextual information more continuously available, technology helps patients to notice otherwise subtle changes and trends in everyday life, which, in turn, increase feelings of safety and comfort \cite{tadas2020barriers,tadas2021transitions}. This aligns with broader motivations reported among adults living with Atrial Fibrillation, whereby self-tracking is often valued as a means to improve awareness and control in daily self-care \cite{keys2025rethinking} and to gain confidence and learn one's bodily limits \cite{tadas2020barriers,tadas2021transitions,andersen_experiences_2020}. In CVD rehabilitation contexts, metrics such as safe training zones can serve as a protective scaffold, supporting individuals who fear exercise while also reducing the risk of overexertion, thereby boosting confidence and motivation to remain active \cite{tadas2020barriers,tadas2021transitions}. Furthermore, during consultations with clinicians, self-tracked data enables anomaly detection, informs lifestyle or medication-related adjustments, and enables more effective communication\cite{keys2025rethinking}. However, self-tracking can be a double-edged sword. While allowing patients to reflect on their health condition, it can also remind them that they are "sick" \cite{ancker_you_2015}. Indeed, (excessive) monitoring could also create anxiety, resulting in higher Atrial Fibrillation symptoms and worse quality of life\cite{rosman_when_2020}. Furthermore, frequent or unclear feedback often triggers anxiety or fatigue \cite{mcmahon2016older,ancker_you_2015,domingos_usability_2022}. To address these, patients valued alerts and warnings most when they were specific and tied to unusual or irregular changes rather than generic, continuous feedback \cite{tadas2021transitions}. 

In the context of heart disease, which significantly affects older adults, there is a need to address their self-tracking and sensemaking needs. Many older adults naturally require daily monitoring beyond clinic visits and rehabilitation sessions \cite{ancker_you_2015,ferguson_wearables_2021}. However, compared with younger adults, older adults are less likely to use digital tools for tracking due to perceived interface complexity and lower comfort with digital technologies \cite{jaana_comparison_2020}. Furthermore, motivation is another barrier, as initial novelty often fades when the presented data lacks personal relevance or actionable meaning \cite{kononova_use_2019}. Thus far, studies on older adults using wearable trackers for self-tracking show both promise and limitations in real-world use \cite{domingos_usability_2022,brickwood_older_2020}. Our study aims to understand how older patients with CVD manage self-tracking in their daily lives and how technology, such as LLMs that support natural-language interaction and a lower technical threshold, can help them in this process. 

\subsection{LLM-based tools in digital health}
LLMs based on natural language processing have recently gained prominence in digital health, supporting tasks ranging from clinical documentation and diagnostic reasoning to patient education, shared decision-making \cite{vs2024towards}, and personalized communication \cite{wang_large_2024}. An increasing number of reviews further highlight this trend \cite{tudor2020conversational,iqbal2025impact,shan2025comparing,takita2025systematic}. LLM-based systems enable self-tracking and sensemaking when integrated with wearable and mobile health apps, capturing rich data streams. Usually embedded as chatbots or voice assistants, these systems are increasingly designed to provide personalized feedback, interpret health data, and support everyday health decisions. Early studies report promising outcomes in health literacy, coaching, data interpretation, and tailored interventions \cite{haag2025last}, as well as in emotional companionship across diverse user groups \cite{jorke2024supporting,munoz_cancerclarity_2025,stromel2024narrating,yang_talk2care_2024}.

By better adapting to users' preferences and personal histories, LLM-based systems have the potential to support self-reflection through narrative feedback and qualitative coaching. For example, GPTCoach applies motivational interviewing techniques to personal health data to guide goal-setting \cite{jorke2024supporting}, while tools such as \textit{CancerClarity} and \textit{NarrativeFitness} interfaces transform quantitative measures into accessible narratives to encourage engagement \cite{munoz_cancerclarity_2025,stromel2024narrating}. 

Research with older adults has emphasized companion robots and community tools that support accessibility, social connection, and general health \cite{enam_artificial_2025,irfan_recommendations_2024}. These works explore relational dynamics, personalization, and interface design, but rarely address independent, self-directed data interpretation. Multimodal systems that link wearable data with LLM feedback for older users remain rare; when tested, they are typically evaluated for technical feasibility rather than lived experience. For instance, Li et al. \cite{li_understanding_2024} developed a multi-sensor system that generated contextual daily summaries with an LLM, but the study focused on integration and performance rather than reflective practices.

At the same time, critiques of LLMs highlight challenges around transparency and trust in healthcare contexts \cite{bjerring_artificial_2021}. While adaptations such as voice input, simplified navigation, and emotional personalization have been suggested \cite{xingyu_li_learning_2025,enam_artificial_2025}, their application to wearable-based self-tracking among older adults with cardiovascular disease remains underexplored. Overall, most such explorations target younger or general adult populations and focus on system capabilities rather than participatory insights into user experience, particularly among older adults with chronic conditions.

\section{Data and Methods}
This section outlines the research design and analysis, recruitment, and participant profile, and study procedure. 

\subsection{Research design and data analysis}

\begin{figure*}[!htb]
\centering
\large
\includegraphics[width=1\linewidth, alt={Overview of the research process}]{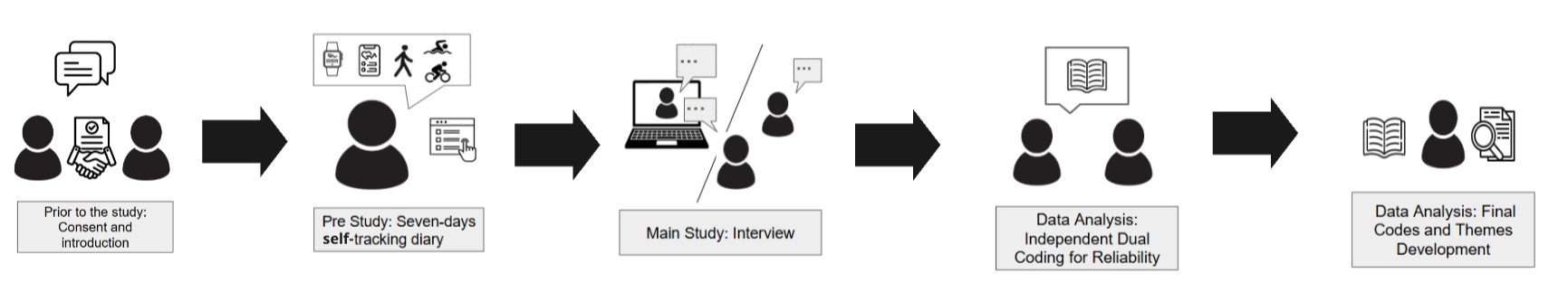}
\caption{Overview of the research process. The study combined a seven-day self-tracking diary with follow-up semi-structured interviews to explore how older adults with CVD engage with self-tracking technologies and interpret their personal health data. Diary entries (text + screenshots) provided contextual insights and informed interview prompts. Interview transcripts formed the primary dataset, which was analyzed using reflexive thematic analysis to develop themes grounded in participants’ lived experiences.}
\label{fig:study-overview}
\end{figure*}

\begin{figure}[hbt!]
    \includegraphics[width=1\linewidth]{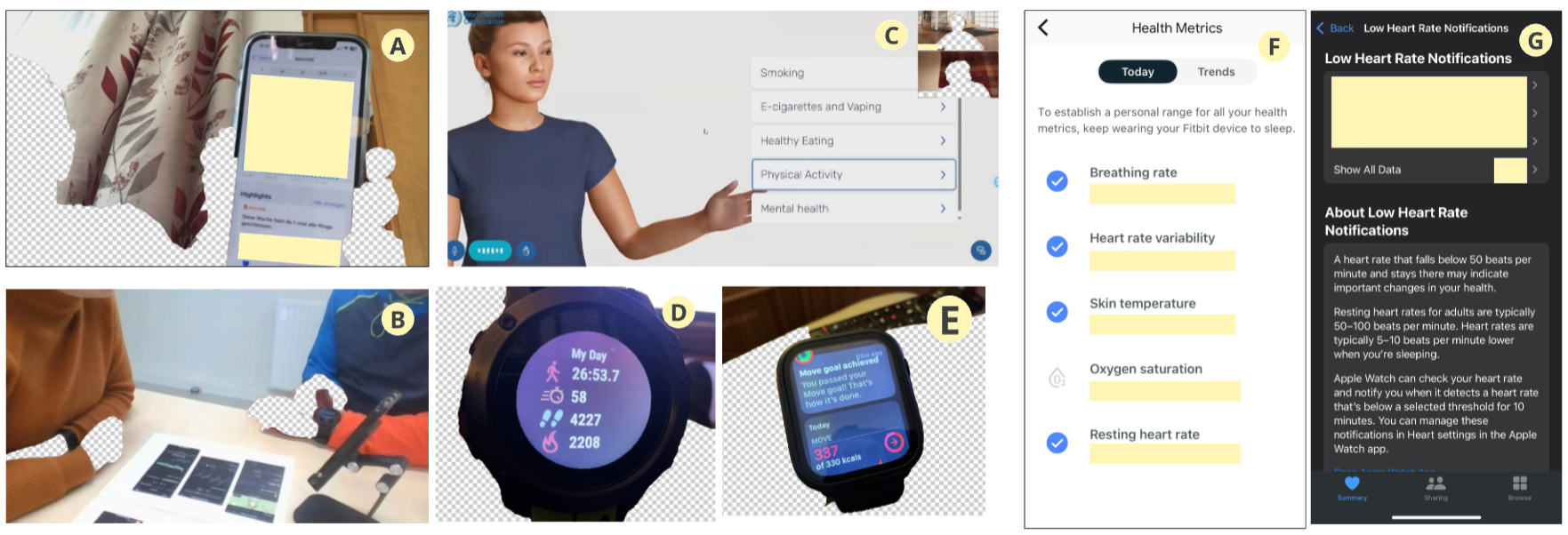}
    \caption{Eight older adults with cardiovascular disease engaged in a seven-day diary study and follow-up interviews. The images represent the study data. Online/In-person interview sessions (A, B, C). Photos uploaded by participants during the 7-day self-tracking diary (D, E, F, G). During one of the online interviews, a participant demonstrated a feature in a mobile health application to illustrate why they found it useful (A). In the in-person interviews, screenshots that participants had submitted during the 7-day self-tracking diary were printed and discussed. Some participants also used their wearable devices alongside the screenshots to elaborate on their experiences (B). As part of the interview procedure, participants were invited to interact with the WHO Health Agent “Sarah” \cite{SarahWHO} and to engage in brief conversations about physical activity knowledge (C). Throughout the 7-day diary study, participants uploaded screenshots from both the wearable interface (D, E) and the mobile health application (F, G). Participants expressed appreciation for the quick daily activity overviews provided by their devices (D) and for motivational feedback on achieving activity goals (E). They also highlighted the value of health metric dashboards for providing a concise overview of key indicators (F), and found anomaly summaries, such as low heart rate notification, particularly useful for monitoring and interpreting potential health changes (G).}
    \label{fig:study-data}
\end{figure}

\begin{figure*}[!htb]
\centering
\large
\includegraphics[width=1\linewidth, alt={Overview of the research process}]{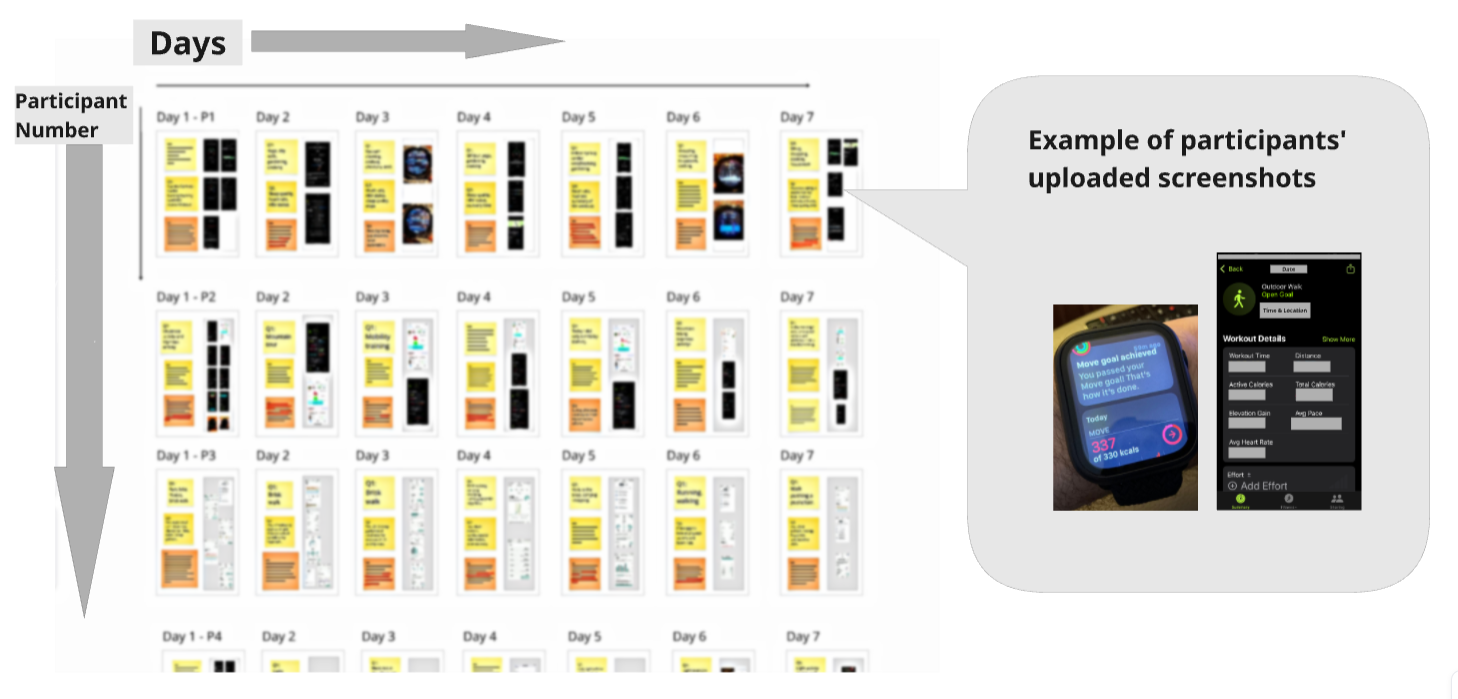}
\caption{The diary study data were organized and summarized using a visual mapping approach on a Miro board (left). The horizontal axis represented the tracking day number (Day 1–7), while the vertical axis represented the participants by participant IDs. Each diary entry was assigned to the corresponding coordinate, yielding a structured overview of responses across time and participants. Examples of participants’ uploaded screenshots are shown on the right.}
\label{fig:diary-data-analysis}
\end{figure*}

This study was a qualitative interview study with eight older adults (n = 8) diagnosed with CVD, investigating how they engage with wearable health technologies and make sense of their health data. Data collection involved two stages (Figure \ref{fig:study-overview}): a seven-day self-tracking diary and individual semi-structured interviews. During the diary phase, participants reflected on their daily routines and device use and submitted screenshots or photos of app interfaces. Inspired by prior work on older adults’ tracking practices through shared images \cite{nurain_i_2023}, this stage focused specifically on health-related tracking. No raw device data (e.g., step counts or heart rate logs) were collected.  \textbf{Diary responses and screenshots informed personalized interview questions}. Figure \ref{fig:study-data} shows the overall interactions with participants and examples of diary data used to prompt semi-structured interviews during the interview study. Figure \ref{fig:diary-data-analysis} illustrates how the diary data were summarized and organized in detail.

Interview transcripts formed the primary basis for analysis. We conducted reflexive thematic analysis~\cite{braun2006using} to identify patterns in participants’ lived experiences while acknowledging the researchers’ interpretive role. To ensure reliability, two researchers independently coded transcripts from three participants chosen randomly from the 8 participants' interview transcripts. Later, another researcher coded the remaining five. In total, 70 codes were identified and grouped into six main themes. Although the sample was small, recurring patterns indicated thematic saturation, consistent with the principle of information power~\cite{malterud2016sample}. Reflexivity was maintained through researcher notes and supervisory feedback.

The qualitative analysis was led by researchers with training in human–computer interaction and prior experience in digital health. While not personally affected by CVD, their expertise in health technologies may have shaped interpretive choices. To mitigate bias, reflexive notes were maintained, and interpretations were regularly discussed within the research team. AI-assisted tools (e.g., ChatGPT) were cautiously explored only for alternative phrasing of theme names, with no participant data shared. 

\subsection{Recruitment and Participants}
Participants were recruited through multiple channels, including Facebook groups on health and aging, local community spaces, a sports center, and outreach by a patient advisory board. Recruitment spanned three countries to increase diversity in CVD experiences and technology use. This multi-country approach enriched the findings by bringing in perspectives shaped by different healthcare and sociocultural contexts, though the small sample size limits cross-country comparisons.

Inclusion criteria required participants to be 60+, have a medically confirmed CVD diagnosis, and have at least 2 months of experience with a wearable device. Initially, English fluency was required, but this was removed after early recruitment challenges to reduce the risk of exclusion. Interested individuals completed a short pre-screening survey.  

In total, eight participants (ages 64–82; 2 Women, 6 Men) took part in the study (Table~\ref{tab:participants}). All were Caucasian and resided in [Blinded for review], [Blinded for review], and [Blinded for review]. They used a range of wearables and health apps and reported regular engagement with their devices, indicating sufficient digital competence for the study. While most lived with partners, one participant lived alone.  

This study received ethical approval from [Blinded for review]. Participants provided written informed consent, could withdraw at any time, and received a [Blinded for review] voucher. Of the approximately 10 initial respondents, one declined participation due to language concerns and another withdrew during data collection.

\begin{table*}[hbt!]
\centering
\caption{Overview of participants. ID = participant number; Gen. = gender; CVD (y) = years since cardiovascular diagnosis.}
\begin{tabular}{|c|l|c|c|c|c|c|}
\hline
\textbf{ID} & \textbf{Location} & \textbf{Age} & \textbf{Gen.} & \textbf{CVD diagnosis (y)} & \textbf{Wearables \& Apps}  \\
\hline
P1 & Salzburg, AT & 64 & M & 6 & Garmin   \\
\hline
P2 & Salzburg, AT & 67 & M & 4 & Garmin   \\
\hline
P3 & London, UK & 66 & F & 12 & Fitbit   \\
\hline
P4 & London, UK & 64 & M & 24 & Apple Health   \\
\hline
P5 & Stockholm, SE & 82 & M & 20 & Apple Health   \\
\hline
P6 & London, UK & 77 & M & 12 & Garmin   \\
\hline
P7 & London, UK & 73 & F & 10 & DaFit   \\
\hline
P8 & London, UK & 70 & M & 10 & Garmin   \\
\hline
\end{tabular}\\[6mm]
\label{tab:participants}
\end{table*}


\subsection{Study Procedure}
Informed consent and onboarding were managed electronically (email/WhatsApp). Given the remote setup, this approach enabled participation across countries but may have introduced a bias toward digitally literate individuals. This was considered acceptable because the study specifically targeted people who already use wearable devices and health apps for self-tracking.

During the seven-day diary phase, participants received daily reminders (5:00 PM, adjusted for one participant’s evening routine) to complete an online questionnaire. Prompts asked them to describe daily activity, reflect on how they accessed and interpreted their health data, and upload screenshots of device interfaces. While most submitted text and screenshots via the form, one participant shared images by email/WhatsApp due to technical issues. Diary entries were then used to guide personalized interview questions.

\begin{figure*}[hbt!]
\centering
\includegraphics[width=1\linewidth, alt={Study Procedure}]{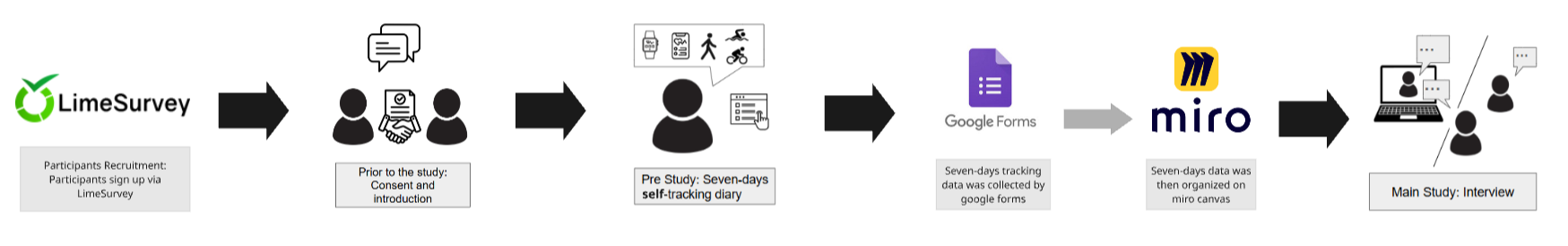}
\caption{Study procedure. Participants were registered electronically and completed a seven-day self-tracking diary with daily reminders. Prompts asked them to reflect on daily activities, describe how they accessed and interpreted health data, and upload screenshots of device interfaces. These diaries were then used to guide one-hour semi-structured interviews, which also included general reflections on tracking and discussion of AI-supported chatbot tools. All interviews were recorded with consent and transcribed verbatim for analysis.}
\label{fig:procedure-overview}
\end{figure*}

Interviews (approx. 1 hour) followed the diary phase. Seven were conducted online (Zoom/Teams) and one in person at [Blinded for review]. They covered: (1) reflections on diary entries and submitted visuals, (2) general tracking practices and challenges, and (3) perspectives on AI-supported health tools. In the final part, participants were introduced to two chatbot examples: a general-purpose health assistant (Sarah, WHO~\cite{SarahWHO}) and a prototype conversational agent for interpreting patient-generated health data in CVD. All interviews were audio- and video-recorded with consent and transcribed verbatim using Whisper~\cite{OpenAIWhisper} (Fig.~\ref{fig:procedure-overview}).


\section{Results}
Six main themes were identified from the interview transcripts. Table~\ref{tab:theme-overview} presents an overview including titles, key insights, and representative quotes. 

\begin{table*}[htbp]
\centering
\caption{Overview of each theme, including title, summary, and representative participant quotes}
\begin{tabularx}{\linewidth}{|p{4cm}| X | X |}
\toprule
\textbf{Theme} & \textbf{Summary}  & \textbf{Representative Quotes} \\
\midrule
Navigating the Emotional Complexity of Self-Tracked Health Data & Health data triggered emotional reactions ranging from anxiety to boredom. This usually depends on data clarity, frequency, and relevance. & \textit{'Some parts in the app do not give me much information, rather it just tells me that there's something wrong with me.'} (P1, Man, 64)
\\
\hline
From Tracking to Making Sense: Owning the Narrative through Self-Tracking & Participants moved from passively following data to confidently interpreting and shaping their own health stories. & \textit{'I'm really experienced and I can explain the data for myself.'} (P2, Man, 67) \\
\hline
Valuing Subjective Experience Over Objective Evidence & Despite using health data, participants relied more on intuition and bodily awareness to guide their actions. & \textit{'If I feel ready, I think I should just go. And if I don’t feel like doing, I’ll just go for a walk.'} (P3, Woman, 66) \\
\hline
Reflecting on Objective Data to Become the Expert on One's Own Body & Participants used data selectively to guide decisions, support self-awareness, and maintain emotional balance in tracking. & \textit{'(Continuing to track physical activity) makes me optimistic about the future... The tracking reinforces my ability to live.'} (P7, Woman, 73)\\
\hline
The Impact of Self-Tracked Data on the Socio-Technical System Surrounding a Patient & Self-tracked data reshaped clinical and social interactions. It enabled patients to prompt more informative dialogue in consultations and share health interpretations socially. & \textit{'The doctor says to me: Don’t push too hard. You can do sports, but always in the safe zone [shown in your tracking tool]...'} (P7, Woman, 73)

 \\
\hline
AI as a Complement for Self-Tracking & Participants imagined AI as future support for interpreting health data and prompting reflection, but emphasized it should not replace personal judgment or clinical advice. & \textit{'It would contribute to making a decision… I can either take that information on board or decide it doesn’t apply to me.'} (P6, Man, 77) \\
\bottomrule
\end{tabularx}
\label{tab:theme-overview}
\end{table*}

\subsection{Theme 1: Navigating the Emotional Complexity of Self-tracked Health Data}
This theme captures the emotional uncertainties participants experienced in their relationship with health data. While self-tracking data created a sense of boredom, anxiety and reassurance depending on the data and their interpretation.

While consistency from the data was considered an indicator for stable health, the unchanging visual representation reduced participants' interest in viewing the data. As P7 said below: 

\begin{quote}
 \textit{'I am always in the right ballpark between 95 and 99 (blood oxygen level). It's a relatively boring screen because it never changes much.'} (P7 , Woman, 73)
\end{quote}

Participants also explained that seeing negative or abnormal results from the health data was not always a bad thing. Examples included high heart rate, low activity, or unusual patterns. They mention that noticing these anomalies helped them become more aware of their health and motivated them to take better care of it. As P7 said:

\begin{quote}
  {\textit{'Because I know what they/(health measurements) normally are for me. So if they're the same as I normally have it, what I'm looking for, I suppose, is something outside the norm.' } (P7, Woman, 73)}

\end{quote}

However, participants also acknowledged  that frequent exposure to negative data or signals of anomalies was mentally exhausting and would usually lead to worry, as P2 noted:

\begin{quote}
    \textit{'It shouldn't be there all the time. It's also like reminding me. Yeah. That's a problem.'} (P2, Man, 67)
\end{quote}

In addition, negative data presented without further explanation made participants feel like there is something wrong with them without providing any reason, as P1 mentioned:

\begin{quote}
  \textit{'Some parts in the app do not give me much information, rather it just tells me that there's something wrong with me.' } (P1, Man, 64)
\end{quote}

Participants generally valued anomalies and changes, but noted a trade-off between receiving too many negative signals without reasoning or reassurance, and the necessity of presenting changes or outliers. 

Some participants adopted strategies to gain a clearer understanding of their health and to approach their data with emotional neutrality, regardless of whether the tracking results were positive or negative. For example, P3 compared herself with peers of a similar age, which provided reassurance:
\begin{quote}
  \textit{'Considering that I am almost 67, I feel that I do quite well compared to my friends. So I think to myself, well, I am actually doing okay.'} (P3, Woman, 66)
\end{quote}

\subsection{Theme 2: From Tracking to Making Sense: Owning the Narrative as CVD Patients}
Living with cardiovascular disease shaped how participants engaged with tracking: their condition gave them both motivation and a strong sense of bodily awareness. Over time, they shifted from passive receivers of numbers to active interpreters who wove data into their personal health narratives. What mattered most was that feedback felt relevant to their condition and respectful of their lived expertise.

Participants interpreted wearable metrics through the lens of their CVD history and embodied experience, using personal benchmarks to judge whether the data was meaningful:
\begin{quote}
\textit{'I just made sure that everything was okay. I wasn't overstressing myself, but I was exercising… People with cardiovascular problems normally have shortness of breath, but I don’t suffer from that. I do get shortness of breath after climbing 200–300 meters, but I can run up and down a flight of stairs without problem.'} (P6, Man, 77)
\end{quote}

Certain metrics, such as heart rate recovery, became symbolic markers of progress and reassurance:  
\begin{quote}
\textit{'The speed it goes from high heart rate back down… that is an indication of how fit I am.'} (P8, Man, 70)
\end{quote}

Several described growing confidence to interpret results without medical input:  
\begin{quote}
\textit{'I’m really experienced and I can explain the data for myself.'} (P2, Man, 67)
\end{quote}

Visualizations in wearables and mHealth apps further supported this self-interpretation and fostered a sense of control, as participants could align the displayed information with their own experiences and conditions: 

\begin{quote}
\textit{'(A bar chart) is easy to understand… it shows you when I’m actually exercising hard.' } (P6, Man, 77)
\end{quote}

For these CVD patients, self-tracking was not just about receiving numbers, but about synthesiszing information elements for constructing and maintaining identity as as knowledgeable, self-managing individuals who balance medical guidance with lived experience.

\subsection{Theme 3: Valuing Subjective Experience Over Objective Evidence}
Participants highlighted how bodily sensations carried more relevance to them than device-generated data. Numbers were seen as useful reference points, but personal experience shaped moment-to-moment decisions about safety, activity, and rest. This reflects their identity as patients who had learned, over years of self-care, to trust embodied knowledge over abstract metrics.

Subjective awareness often guided choices even when devices reported differently. One participant recalled:  
\begin{quote}
\textit{'If I feel ready, I think I should just go. And if I don’t feel like doing, I’ll just do a walk.' } (P3, Woman, 66)
\end{quote}

Wearables were valued more as reassurance than directive tools. During an episode of breathlessness, P7 checked her device and decided not to seek further medical help:
\begin{quote}
\textit{'My oxygen level was still high… so I thought, no, I’m just going to sit and relax.' } (P7, Woman, 73)
\end{quote}

Participants described data as incomplete without context. For example, P3 explained that physical symptoms of bradycardia, which was the slowing of heart rate, mattered more than temporary readings:  
\begin{quote}
\textit{'I don’t usually go to the doctor right away when the health metrics on the wearables are not normal… if it continues for two or three days, then I get physical symptoms of bradycardia.'} (P3, Woman, 66)
\end{quote}

When device metrics matched participants’ bodily sensations, the alignment reinforced their trust and confidence in both themselves and the device. For instance, one participant described how their smart watch showed a normal atrial fibrillation reading, which corresponded with their own feeling of being symptom-free:  

\begin{quote}
\textit{'The watch shows my atrial fibrillation as normal, that is the lowest level it can register — and I also rely on my own feelings to confirm the result, since I haven’t had any problems with it.' } (P5, Man, 82)
\end{quote}

Overall, this tendency to first listen to their bodies reflects an important stance: as CVD patients, participants took responsibility for their own health, treating devices as companions that validated their embodied sense of wellbeing rather than replacing their ability to care for themselves.

\subsection{Theme 4: Reflecting on Objective Data to Become the Expert of One's Own Body}
Rather than relying solely on medical professionals or digital systems, participants took a more active role in managing their health by interpreting self-tracking data in ways that aligned with their personal context. Their engagement with tracking was intentional, selective and grounded in a desire for autonomy. The question '\textit{How can we become experts on our own body?}' guided much of their behavior.

Participants selectively chose to engage with information that was useful for both their own understanding and to share with their doctors. 
\begin{quote}
\textit{'I don't want to look at it so often so it'll tell me by kind of exclusion. Maybe send a message to the cardiologist to say that it's seen something. Have a look at that and then it messages me back. you can switch off to it and it'll only tell you if it's concerned itself.'} (P4, Man, 64)
\end{quote}

For some, tracking contributed to a feeling of optimism and security. The ability to observe and respond to bodily patterns reinforced their belief that they were actively doing something to protect their well-being:

\begin{quote}
\textit{'(Continuing to track physical activity) makes me optimistic about the future. I thought I might die of another heart attack sooner rather than later. The tracking reinforces my ability to live.'}  (P7, Woman , 73)
\end{quote}

Self-tracking may provide participants with an external, objective perspective on their bodies, prompting them to reflect more deeply on their condition. 
\begin{quote}
\textit{'I may have a different point of view if I was chronically with my heart very bad. If I was having to see the doctor instead of once a year, I was having to see him once a month. More information would be more relevant and important.'} (P6 , 77, Man)
\end{quote}

\subsection{Theme 5: Self-tracked data can reshape the socio-technical system surrounding a patient}

For participants living with CVD, managing health was closely tied to bodily safety, symptom interpretation, and long-term self-care. Wearable and mobile health technologies enabled them to track key metrics such as heart rate, blood oxygen, and ECG values (relevant to atrial fibrillation), which they actively interpreted and brought into conversations with others. These interpretations were not confined to individual reflection but became part of broader dialogues, both within clinical consultations and in everyday social interactions.

\paragraph{Data-driven dialogues in cardiac consultations}

Participants brought self-tracked data into consultations to prompt medication adjustments or clarify anomalies, shifting interactions from unquestioned authority to dialogue grounded in clinical expertise and lived experience. For example, one participant initiated a discussion after their pulse repeatedly exceeded the recommended range during exercise:

\begin{quote}
    \textit{'Because the cardiac rehab that I had said my pulse shouldn't go above a certain level and it was going above that level every time so I thought I'll go and see a cardiologist who said well if you feel okay and you've run for years don’t worry so that was really useful.'} (P3,Woman,66)
\end{quote}

The use of self-tracked data not only helped patients scaffold medical dialogues during consultations but also influenced providers’ expectations of what could be monitored and shared. For example, one participant recalled being advised to remain within the ‘safe zone’ displayed on her device:

\begin{quote}
\textit{'The doctor says to me: Don’t push too hard. You can do sports, but always in the safe zone [shown in your tracking tool]...'} (P7, Woman, 73)
\end{quote}

In other cases, patient-generated data directly informed treatment decisions:

\begin{quote}
\textit{'We took my blood pressure morning and night, and then my husband printed it out, took it to the consultation... on that basis, he (healthcare provider) altered the drugs.' }(P7, Woman, 73)
\end{quote}

Thus, self-tracking reshaped clinical consultation, enhancing patients’ bodily awareness, supporting shared decision-making, and raising providers’ expectations for continuous monitoring.

\paragraph{Dialogues with Social Connections}

Participants also engaged in informal sensemaking through everyday conversations, where self-tracked data became a shared point of reference. Some compared their results with friends or training partners, which helped contextualize their own readings and turned data reflection into part of everyday social interaction:  

\begin{quote}
    \textit{'When I was on some of the machines my friend goes with us and we may be walking and after half a mile his heart rate is like up to 100 and I’m still 70 or 80.'} (P4, Man, 64)

   \textit{'Yes, I discussed with friends about how my training was and so on.' }(P1, Man, 64)
\end{quote}

Others described how self-tracked data fostered interaction with those around them, whether to seek clarification, engage in everyday conversation, or share positive feedback as a source of encouragement and joy:

\begin{quote}
    \textit{'If something looks strange, I’ll ask my daughter, since she’s a doctor. My son-in-law is also quite into fitness and has a tracker, so I’ll chat with him about my results. And with my husband, I’ll say, ‘oh look at this,’ when the feedback is good.' } (P3, Woman, 66)
\end{quote}

Overall, self-tracking CVD patients sought reassurance and safety not only through their own interpretations but also through data-driven dialogues with doctors and social connections.

\subsection{Theme 6: AI as a Complement for Self-Tracking}
While social interactions were central to participants’ current practices, AI-supported tools were discussed as hypothetical, future-oriented companions. Participants reflected on three imagined AI functionalities: (1) sensemaking of health data, (2) suggesting exercise adjustments, and (3) encouraging reflection or more frequent tracking.

Many participants expressed cautious optimism, seeing potential for AI to help them interpret patterns, provide personalized insights, or communicate findings to healthcare providers, as long as it complemented rather than replaced human judgment:
\begin{quote}
\textit{'I would like to think about it as a Jiminy Cricket… it sits on your shoulder and gives advice.' } (P4, Man, 64)
\end{quote}

Others valued the accessibility of chat-based summaries that could be revisited later, though trust in AI-generated recommendations was often conditional. Participants stressed the need for expert oversight and personalized context:
\begin{quote}
\textit{'If you feel that the information is true, you will implement it... but if [AI] does not know you well enough, it might provide inaccurate information.' }(P1, Man, 64)
\end{quote}

AI recommendations were therefore seen as supportive rather than decisive inputs:
\begin{quote}
\textit{'It would contribute to making a decision… I can either take that information on board or decide it doesn’t apply to me.' } (P6, Man, 77)
\end{quote}

Concerns included information overload and anxiety, with some recalling disabling automated summaries when they became repetitive. Motivational reminders were met with mixed responses—light, goal-based encouragement was welcomed, but frequent nudges felt intrusive:
\begin{quote}
\textit{'I didn't like that... it's not natural. But if it says you've done 13,000 steps today, your goal was 10,000, that motivates me.' }(P3, Woman, 66)
\end{quote}

Overall, participants envisioned AI as a complementary support. AI was perceived as a mediator that could enhance sensemaking and confidence in patient–provider interactions, without undermining embodied awareness or clinical authority.

\section{Discussion}
The discussion builds on the six key themes summarized in the results and relates them to existing literature. These findings address both \textbf{RQ1} and \textbf{RQ2}. \textbf{RQ1 }is discussed in \textbf{Section 5.1}. This section synthesizes the results on strategies, challenges, and motivations that older adults with CVD encounter in their self-tracking practices. \textbf{RQ2} is discussed in \textbf{Section 5.2}. This section focuses on how LLM-based chatbots can support wearable data sensemaking among older adults with CVD. The design directions are summarized in Figure \ref{tab:direction-overview} in the appendix. They include designing for 1) emotional engagement, 2) supporting personal agency, 3) enabling data-guided selective engagement, 4) acknowledging embodied experiences, 5) prompting collaborative conversations, and 6) providing respectful agent support. 

\subsection{RQ1: Strategies, Barriers, and Motivations in Self-Tracking Practices of Older Adults with CVD}
Older adults with CVD engaged in self-tracking through a combination of personal strategies, social practices, and emotional negotiations. Many participants focused on values they found meaningful, such as step count or heart rate recovery, while ignoring indicators that felt irrelevant or too detailed.  Bodily awareness also shaped interpretation. This selective interpretation helped reduce cognitive burden and supported daily use \cite{wienroth_health_2020,wang_redefining_2024}. Participants often trusted their own feelings more than device outputs, using numbers only as a reference rather than as the main guide to decision-making \cite{ancker_you_2015,nurain_i_2023,vargemidis_performance_2023,kononova_use_2019}. Beyond individual practices, self-tracked data also played a social role. Participants brought their records into clinical consultations to clarify inconsistencies, support treatment decisions, and take a more active part in care \cite{augst_patientgenerated_2025,zhu2017sharing}. Data was also shared with family and peers to validate experiences, receive encouragement, and build intimacy \cite{lupton2021sharing,Rooksby2014LivedInformatics,Kaziunas2017CaringData}.

At the same time, participants encountered several barriers. Unclear or negative outputs often created anxiety, confusion, or fatigue, and in some cases led to disengagement from tracking \cite{ancker_you_2015,Kaziunas2017CaringData,andersen_experiences_2020}. A strong sense of agency also carried risks. Many relied on their own judgment, often shaped by experiences in cardiac rehabilitation, which gave them confidence in interpreting data but could lead to confirmation bias or misinterpretation \cite{wang_redefining_2024,rooksby_personal_2014,meppelink_i_2019}. Barriers also arose in clinical settings when patient-generated data was questioned or dismissed. Such moments undermined participants’ sense of contribution and created tension in collaborative care \cite{augst_patientgenerated_2025}.

Despite these challenges, participants described several motivations for self-tracking. Positive results often brought reassurance and comfort, while anomalies encouraged reflection and sometimes prompted dialogue with healthcare providers \cite{ancker_you_2015,andersen_experiences_2020,Kaziunas2017CaringData}. Self-tracked data was also valued as a preparation tool that helped participants feel more informed and confident before medical encounters \cite{zhu2017sharing,augst_patientgenerated_2025}. Sharing data with family and peers added further motivation by providing support, encouragement, and emotional connection \cite{lupton2021sharing,Rooksby2014LivedInformatics}. 

\subsection{RQ2: Designing LLM-enabled health data Sensemaking Tools}
older adults managing chronic conditions such as CVD often have more complex and ongoing sensemaking needs than the general adult population. They tend to engage more actively in self-tracking and frequently seek reassurance when data appears unclear or unexpected \cite{Keys2024}. LLM-based chatbots are well suited to this context because they enable natural, adaptive dialogue. Prior studies show that older adults value technologies that provide both clarity and emotional responsiveness \cite{enam_artificial_2025, wang_redefining_2024, vargemidis_performance_2023}. Chat-based interfaces can deliver these qualities by offering personalized feedback in plain language while also responding to emotional cues. They build on familiar patterns of everyday conversation, which can feel more intuitive than technical dashboards or abstract data visualizations. In doing so, conversational interfaces may foster not only clearer sensemaking but also a greater sense of engagement and companionship. In our study, while participants expressed cautious optimism toward AI, they appreciated tools that could help analyze patterns, highlight relevant metrics, and provide natural language feedback, as long as these systems acted as supportive collaborators rather than authoritative voices \cite{stromel2024narrating,li_stage-based_2010,wienroth_health_2020,kim_how_2024}.

\subsubsection{Designing for Emotional Engagement.}
This study supports existing research emphasizing the complex emotional relationship older adult CVD patients have with self-tracking technologies, especially when interpreting health-related data \cite{ancker_you_2015,Kaziunas2017CaringData,andersen_experiences_2020}. This is clearly reflected in Theme 1:\textit{Navigating the Emotional Complexity in Self-tracked Health Data}, where participants described a wide range of emotional responses including reassurance, confusion, worry and emotional fatigue. Kinesiophobia -- the fear of movement physical activity -- is common amongst CVD patients \cite{sahin_effect_2021}. As such embedding emotional support could facilitate the data-sensemaking process.

To address this, LLM-based chatbots could detect emotional cues like “I feel worried about my Heart rate” and respond with calm and simple explanations. While the chatbot’s primary role in this context is reflection and emotional support rather than diagnosis, patient safety still requires that even supportive systems are rigorously tested \cite{kim_how_2024}. This ensures they do not accidentally mislead users, overstep into clinical advice, or create unsafe guidance. Clear boundaries must therefore be maintained: the chatbot may reassure, invite reflection, and hold space for emotions, but it has to be placed carefully to not attempt to diagnose or prescribe. For older adults in particular, such boundaries help reduce confusion while still fostering engagement and trust. When emotional signals are recognized, the system might follow up with gentle prompts like “Does this result make you feel anxious or unsure? Would you like to explore a bit why together?”. Such conversation turns negative reactions into opportunities for deeper understanding and reassurance, while preserving autonomy \cite{tao_effects_2019, timmermans_presenting_2008, tadas_using_2023}.

\begin{designimp}[Design Direction: Designing for Emotional Engagement]
LLM-based chatbots could detect emotional cues like confusion and respond with simple, reassuring interpretations that acknowledge older CVD patients concerns while avoid making diagnosis or medical decisions. 
\end{designimp}

\subsubsection{Designing for Improved Agency and Empowerment.} 
Older adults’ engagement with self-tracking depends not only on accuracy but on \textbf{the control they feel in interpreting data} \cite{wang_redefining_2024,rooksby_personal_2014}. Prior work shows that agency grows when users are allowed to make meaning themselves rather than being directed \cite{bjerring_artificial_2021,kim_how_2024}. Based on the insights, LLM-based chatbots could \textbf{act as guides, not authorities}. They can scaffold reflection with prompts that broaden reasoning without overtaking decision-making. For example: “Would you like to explore whether this relates to your sleep or activity patterns?” \cite{jorke2024supporting,mamykina_adopting_2015,coskun_data_2023}. Many participants showed \textbf{confidence rooted in cardiac rehabilitation}, where structured monitoring improved health literacy. While empowering, such confidence risks confirmation bias \cite{meppelink_i_2019}. The chatbots could be designed to respect user reasoning while gently introducing alternatives, e.g., “You usually link low activity to poor sleep. Would you like to consider other factors too?” 

\begin{designimp}[Design Direction: Designing for Improved Agency and Empowerment]
Agency is central for older adults interpreting their own health data. LLM-based chatbots can support this by adapting to individual reasoning patterns and offering prompts that expand reflection without taking over decision-making. Yet, confidence in self-interpretation can slip into overconfidence, leading to confirmation bias. To safeguard against this, systems could integrate expert-in-the-loop mechanisms \cite{ramjee_cataractbot_2025}, where uncertain cases are escalated for asynchronous clinical feedback. This balance preserves agency while ensuring safety. 
\end{designimp}

\subsubsection{Designing for Embodied Acknowledgment.}
Beyond guiding data interpretation, LLM-based chatbots will need to account for how older adults experience their health through the body. Our findings show that participants often \textbf{prioritized subjective sensations} over device readings, echoing prior work on embodied health sensemaking \cite{ancker_you_2015,nurain_i_2023,vargemidis_performance_2023,kononova_use_2019}. Rather than dismissing these reports, systems should validate and incorporate them into feedback, supporting older adults’ trust in their own awareness while contextualizing the data \cite{benbunan2019affordance,vargemidis_irrelevant_2021}. This design direction addresses a known tension: older adults with CVD navigate between \textbf{trusting bodily sensations} and \textbf{relying on wearable metrics}. Rely too much on devices can heighten anxiety when alerts feel incongruent with the body \cite{rosman_wearable_2024}, while dismissing critical alerts may cause users to miss early signs of worsening health, such as irregular heart rhythms or increased fatigue \cite{andersen_experiences_2020}. Chatbots could mediate this balance by not only acknowledging embodied experience, but also highlighting anomalies and their potential implication for getting medical support \cite{guasti_digital_2022}.

\begin{designimp}[Design Direction: Designing for Embodied Acknowledgment]
LLM-based chatbots could recognize and contextualize users’ bodily sensations alongside device metrics, balancing subjective experience with clinical signals to reduce anxiety, prevent missed health warnings, and support a more holistic sensemaking process.  
\end{designimp}

\subsubsection{Designing for Data-Driven Guidance.}
LLMs are powerful tools for health sensemaking since they can analyze patterns across large data streams and answer natural language queries \cite{stromel2024narrating}. Yet our findings show that older adults engage selectively with tracking data: they attend to metrics that feel relevant in daily life and ignore those that seem irrelevant or overwhelming. This selective approach reflects both a desire for control and a strategy to reduce cognitive burden, echoing prior work on how older adults actively construct meaning from their data \cite{wang_redefining_2024, wienroth_health_2020}.

To support these practices, LLM-based chatbots should prioritize the metrics users most often consult (e.g., step counts or resting heart rate) to reduce overload while maintaining clarity. At the same time, systems must preserve transparency: older adults need to understand why an insight has changed, especially when interpretations differ from their prior expectations \cite{de_thurah_understanding_2025}. Personalization is therefore less about the model itself than about surrounding interaction design. How preferences are tracked, how explanations are presented, and how interfaces balance conversational and rule-based elements \cite{jaana_comparison_2020, vivion_how_2024}.

This balance is critical for older adults with CVD. Many participants displayed strong confidence in interpreting their own metrics, but selective attention risks overlooking early signs of worsening health \cite{hu_physical_2022}. Systems could therefore respect user-defined focus while also adapting to medical priorities, surfacing anomalies or critical changes when necessary. In this way, LLM-based chatbots can reinforce agency without allowing important signals to go unnoticed. To safeguard against errors, \textbf{expert-in-the-loop mechanisms are crucial}. Following Ramjee et al. \cite{ramjee_cataractbot_2025}, uncertain queries could be flagged for \textbf{asynchronous clinician review}. This reduces burden on providers while ensuring accuracy and safety, with feedback also improving models over time. They receive accurate, unbiased information is paramount for \textbf{patient safety}.

\begin{designimp}[Design Direction: Designing for Data-Driven Guidance]
LLM-based chatbots could align with older adults’ selective engagement by prioritizing familiar metrics, while still surfacing critical anomalies to avoid missing health risks. Safety could be supported through expert-in-the-loop triage: if anomalies persist, fall outside baselines, or users express worry, the system generates a plain-language summary and offers options such as “queue for clinician review” or “save for next visit.” Boundaries remain clear—never diagnose, always communicate uncertainty, follow provider guidance, and preserve user control over what is shared.
\end{designimp}

\subsubsection{Designing for Collaborative Dialogue.}
From theme 5, we knew that older CVD patients used self-tracked data to initiate dialogue with healthcare providers and peers. While it enables patients to make their experiences heard, tension arises when data is questioned or dismissed. Designing AI tools like LLM-based chatbots could consider not only what information is presented, but also how it is framed and interpreted in social and clinical contexts. 

To support evolving and collaborative relationships in CVD care, LLM-based chatbots could serve as conversational tools that help patients prepare for and engage in meaningful health dialogues. For example, the chatbot could automatically generate printable summaries of weekly trends such as fluctuation in heart rate or step counts, in natural language that patients can bring to clinical appointments. However, self-tracked data is not just clinical evidence but also emotionally charged. The feature such as printable report might share everything, even data patients feel uncomfortable to share with. This raises design and ethical challenges: chatbots must respect privacy, support selective data sharing, and scaffold patients agency over what is shared, when and with whom \cite{wieczorek2023ethics}.

\begin{designimp}[Design Direction: Designing for Collaborative Dialogue]
LLM-based chatbots could support collaborative dialogue by helping older CVD patients prepare for appointments, summarize key patterns, and decide what to share. To avoid ethical pitfalls, these tools should be designed carefully by enabling selective sharing and respecting privacy. 
\end{designimp}

\subsubsection{Designing for Respectful Agent Support.}
Respectful agent support begins with context-aware and conversational insight generation. Participants valued chatbots that highlighted trends in a supportive, non-judgmental tone, helping them reflect without prescribing actions. This aligns with Li et al.’s Stage-Based Model of Personal Informatics \cite{li_stage-based_2010}, where reflection fosters meaning-making without imposing behavior change.

Trust hinged less on technical capability than on how the chatbot positioned itself. Feedback framed as questions or reflections was seen as credible and autonomy-preserving, whereas prescriptive prompts risked undermining trust \cite{kim_how_2024,stromel2024narrating,li_stage-based_2010,wienroth_health_2020}. Echoing Kim et al. \cite{kim_how_2024}, participants warned that behavioral suggestions misaligned with provider advice reduced chatbots to redundant intermediaries. Since self-tracked data also carries social and emotional weight, relational sensitivity is critical for sustaining trust \cite{wieczorek2023ethics}.

\begin{designimp}[Design Direction: Designing for Respectful Agent Support]
LLM-based chatbots need to be designed to provide context-aware, sensitive personalized conversations in a supportive, non-judgmental tone. Their role need to be carefully positioned as a helpful complement between patient and provider, rather than a redundant intermediary.
\end{designimp}

Our exploratory findings can motivate other works beyond CVD, contributing to broader works around chronic / non-communicable disease, data sensemaking for health (behavior) support and HCI discussions of personal informatics and human–data interaction. The themes we identify, namely emotion-aware reframing, embodied interpretation, and selective engagement, resonate with foundational work on lived data practices and legibility in personal informatics \cite{li_stage-based_2010,rooksby_personal_2014}. \textbf{However, we acknowledge that given the risk of LLM misuse \cite{blum_misuse_2026} patient-safety is a key consideration while allowing patients to independently engage with their data. There is a need to ensure that the LLM's do not give false reassurances to patients leading to delayed care. Therefore it is essential for having oversight over interactions patients might have with LLM tools -- while respecting patients' privacy boundaries -- and access to medical support in case of uncertainty in results.}

By articulating design principles for LLM-enabled systems, we bridge to ongoing efforts in behavior change technologies \cite{consolvo2009theory} and negotiation of data meaning in socio-technical ecologies \cite{elsden2016quantified}, offering motivation for future work investigating older-adult self-tracking as well as chronic disease management more broadly.

\subsection{Future Work}
Future research should focus on designing and evaluating LLM-based technologies integrated with real-time feedback from wearable data with CVD patients. Such systems could test how personalized, emotionally supportive feedback helps older adults interpret patterns and anomalies in reflective ways, while avoiding causing stress or discouragement. Participatory co-design workshops are also needed to involve older adults with CVD in shaping interaction styles, preferred language, and features that foster a sense of control. Designing AI with, rather than for, older adults ensures alignment with their values and daily routines. Long-term studies should further examine how trust, motivation, and use evolve over time, and whether AI-based sensemaking tools influence behavior change, reduce emotional strain, or improve communication with healthcare providers.

\subsection{Study Limitations}
This study provides insight into how older adults with CVD engage in self-tracking and envision AI-based support, but several limitations remain. The sample was small (eight participants), homogenous in ethnicity, and relatively healthy, limiting generalizability. Participants were not recruited by condition severity, which meant findings often reflected general self-tracking practices rather than CVD-specific challenges. Additionally, the data collected in the study represent participants' perceptions on self-tracking technologies, which indicates that more research on their actual use patterns with LLM-based technologies is required. Nonetheless, the study captured perspectives of older adults with CVD often underrepresented in research, offering valuable formative insights into how this group interprets self-tracked data and imagines intelligent support.  

\section{Conclusion}
This study explored how older adults with CVD engage with self-tracked health data and how LLM-based systems might support sensemaking. Participants showed a strong interest in reflecting on and personalizing their data, often reinterpreting metrics through bodily awareness and daily routines rather than relying solely on technology or clinicians. 
They were open to AI-guided support provided as non-directive, adaptive, and respectful of their lived context. They envisioned it as a companion that prompts reflection rather than offering prescriptive advice. Self-tracked data also played a key role in clinical and social dialogues, underscoring that data sensemaking is not only individual but social and affective process. Overall, the findings suggest that designing for emotional awareness, agency, guidance, and embodied acknowledgment in LLMs, with safety as a key consideration, can make digital health sensemaking tools more meaningful and supportive for older adults managing chronic illnesses, especially CVD.





\bibliographystyle{ACM-Reference-Format}
\bibliography{sample-base}

\appendix
\begin{table*}[htbp]
\centering
\caption{Overview of LLM-based chat bot design directions for older adults and older adults with CVD diagnosis}
\begin{tabularx}{\linewidth}{|p{4cm}| X | X |}
\toprule
 & Designing for Older Adults  & Designing for older adults with CVD diagnosis \\
\midrule
 Designing for Emotional Engagement& \begin{minipage}[t]{\linewidth}
\begin{itemize}
    \item Recognize emotional cues related to health data.
    \item Provide simple explanations or possible reasons for unclear or confusing data.
\end{itemize}
\end{minipage} & Continue to recognize emotional cues and explain data clearly. The chat bot should stay calm and not overreact. It should respond with care, like a supportive nurse, not act like a doctor. \\
\hline
Designing for Improved Agency and Empowerment & \begin{minipage}[t]{\linewidth}
\begin{itemize}
    \item Support independent data interpretation through guided dialogue and reference to previous tracking history.
    \item Clearly explain updates to maintain understanding.
\end{itemize}
\end{minipage} & In addition, avoid reinforcing overconfidence and confirmation bias. Encourage critical reflection on personal interpretations and guide users to reassess assumptions when needed. \\
\hline
 Designing for Data-Driven Guidance & \begin{minipage}[t]{\linewidth}
\begin{itemize}
    \item Highlight relevant metrics and simplify trends to reduce cognitive overload.
    \item Personalize data presentation based on individual tracking preferences.
      \item Consider Expert-In-The Loop to ensure patient safety
\end{itemize}
\end{minipage} & In addition to supporting selective tracking, ensure critical CVD-related metrics are monitored. Prompt users when anomalies appear, even if the data is not frequently viewed. \\
\hline
Designing for Embodied Acknowledgment & \begin{minipage}[t]{\linewidth}
\begin{itemize}
    \item Acknowledge and contextualize subjective feelings of older adults, even when the data appears normal. Support reflection by helping them explore mismatches between bodily experience and tracking data.
\end{itemize}
\end{minipage} & In addition to validating subjective feelings, the chatbot should help older adults balance bodily cues with objective data, especially when symptoms are silent. Encourage careful, calm sensemaking without causing unnecessary alarm. \\
\hline
Designing for Collaborative Dialogue & 
\begin{minipage}[t]{\linewidth}
\begin{itemize}
    \item Help users reflect on trends and generate summaries they can share in clinical conversations.
    \item Use prompts that encourage older adults to prompt social conversations with their family or peers.
\end{itemize}
\end{minipage} & 
Use prompts that encourage patients to prepare questions or observations for doctors. Ensure the chatbot acts as a bridge between patient and provider, not a replacement. \\
\hline
Designing for Respectful Agent Support & 
\begin{minipage}[t]{\linewidth}
\begin{itemize}
    \item Use ask-not-tell language for behavior prompts.
    \item Allow older adults to customize the frequency, tone, and timing of feedback.
\end{itemize}
\end{minipage} & 
Avoid unsolicited or overly directive suggestions. Instead, offer optional, context-sensitive guidance that supports autonomy, aligns with medical advice, and avoids redundancy or confusion. \\
\bottomrule
\end{tabularx}
\label{tab:direction-overview}
\end{table*}
\end{document}